\documentclass[10pt]{iopart}

\usepackage[utf8]{inputenc}
\usepackage{graphicx}
\usepackage{xcolor}
\usepackage{soul}

\begin{document}

\title[]{Nonlinear MHD modeling of soft $\beta$ limits in W7-AS}

\author{R Ramasamy$^{1}$, K Aleynikova$^2$, N Nikulsin$^3$, F Hindenlang$^1$, I Holod$^1$, E Strumberger$^1$, M Hoelzl$^1$ and the JOREK team$^4$}
\address{$^1$ Max-Planck Institut f\"ur Plasmaphysik, Boltzmannstraße 2, 85748 Garching bei München, Germany}
\address{$^2$ Max-Planck Institut f\"ur Plasmaphysik, Wendelsteinstrasse 1, 17491 Greifswald, Germany}
\address{$^3$ Dept. of Astrophysical Sciences, Princeton University}
\address{$^4$ see list of contributors in \cite{hoelzl2021jorek}} 









\begin{abstract}
      An important question for the outlook of stellarator reactors is their robustness against pressure driven modes, and the underlying mechanism behind experimentally observed soft $\beta$ limits. Towards building a robust answer to these questions, simulation studies are presented using a recently derived reduced nonlinear MHD model. First, the initial model implementation is extended to capture fluid compression by including the influence of parallel flows. Linear benchmarks of a (2, 1) tearing mode in W7-AS geometry, and interchange modes in a finite $\beta$, net-zero current carrying stellarator with low magnetic shear are then used to demonstrate the modeling capabilities. Finally, a validation study is conducted on experimental reconstructions of finite $\beta$ W7-AS discharges. In agreement with past experimental analysis, it is shown that (i) the MHD activity is resistive, (ii) a soft $\beta$ limit is observed, when the plasma resistivity approaches the estimated experimental value, and (iii) low $n$ MHD activity is observed at intermediate $\beta$ values, particularly a nonlinearly dominant (2, 1) mode. The MHD activity is mild, explaining the soft $\beta$ limit, because the plasma volume remains separated into distinct sub-volumes in which field lines are ergodically confined. For the assumed transport parameters, the enhanced perpendicular transport along stochastic magnetic field lines can be overcome with the experimental heating power. The limitations in the current modeling are described, alongside an outlook for characterising soft $\beta$ limits in more detail in future work.
\end{abstract}



\submitto{\NF}
\maketitle
\ioptwocol

\section{Introduction} \label{sec:intro}
Cross experimental databases support the conclusion that the $\beta$ limit of net zero current carrying stellarators are often soft, in the sense that large scale disruptive magnetohydrodynamic (MHD) activity is avoided \cite{weller2006significance}. As a result, the operational limit prescribed by Mercier's criterion can be exceeded experimentally \cite{weller2001, de2015magnetic, watanabe2005effects}. This indicates that, unlike in tokamaks, linear MHD stability thresholds overconstrain the solution space for future viable stellarator reactors. In such a way, the fusion community needs to seek out an improved method of bounding the solution space in stellarator optimisation of high $\beta$ designs, incorporating nonlinear effects.

Towards this end, a better understanding of the underlying mechanism behind the nonlinear saturation of pressure driven modes in stellarators is necessary. The immediate trouble with this is that the manifestation of so-called soft $\beta$ limits is somewhat broad. The behaviour of Wendelstein 7-AS (W7-AS) is characterised by its partial optimisation, in the sense that increasing Pfirsch-Schl\"uter currents with increasing plasma $\beta$ lead to a deepening of the magnetic well and tokamak-like shear, stabilising pressure driven modes. This phenomenon is very different from the super dense core (SDC) regime observed in the Large Helical Device (LHD). In this operational scenario, the rotational transform crosses the $\iota=1$ rational surface, and the magnetic well is intentionally weakened at high $\beta$ in order to improve particle and heat confinement \cite{ohdachi2017observation}. Based on these experimental data points, it is difficult to predict the detailed nonlinear behaviour of Wendelstein 7-X (W7-X), which is designed with a vacuum magnetic well, and low magnetic shear even at high $\beta$ due to its neoclassical optimisation. The same can be said for quasi-symmetric (QS) devices, where a significant bootstrap current contributes to the confining magnetic field, complicating the dynamics. 

For this reason, it is important to develop nonlinear numerical methods that can correctly reproduce and interrogate the behaviour of existing and future experimental devices. Several of the main workhorse nonlinear codes have either demonstrated recent nonlinear MHD extensions for stellarators, or are in the process of being extended for such studies \cite{zhou2021approach, sovinec2021initial, nikulsin2022jorek3d}. In addition, the MIPS code has been used extensively for nonlinear MHD studies of stellarators \cite{suzuki2021nonlinear}. While such a detailed nonlinear approach cannot be applied to stellarator optimisation directly, it can help to inform and direct the development of a nonlinear analytic approach. Further, beyond the task of stellarator optimisation, a high fidelity model of the detailed reactor performance that such codes can provide will be valuable for understanding MHD phenomena in stellarators.


The current work presents recent developments of the JOREK code \cite{hoelzl2021jorek, huysmans2009non} towards these goals. In previous work, a hierarchy of stellarator capable reduced and full MHD models was derived, and the simplest of these models was implemented and tested in the JOREK code \cite{nikulsin2022jorek3d, nikulsin2021models}. This contribution covers the latest developments of the code towards assessments of finite $\beta$ effects, by extending the model in JOREK to include parallel flows, and applying the model to more advanced cases than previously considered. Herein, $\beta$ is used to refer to the volume averaged ratio of the plasma fluid  pressure to magnetic pressure.

The remainder of the paper is outlined as follows. In Section \ref{sec:reduced_mhd}, the reduced MHD model implemented in JOREK is discussed. In particular, the extension of the original model implementation to include parallel flow dynamics, and the potential danger posed by neglecting compressional Alfv\'enic activity in reduced MHD models for stellarators are covered. Recent analytical work suggests that the influence of the fast magnetosonic wave can be important in regions with negligible magnetic shear \cite{graves2019reduced}. This regime is close to the operational space of many stellarator experiments, which can have weak magnetic shear, and so the analytical result needs to be taken seriously by any reduced MHD approach.

Verification studies are carried out and used to justify the use of reduced MHD to study soft $\beta$ limits in Section \ref{sec:verification}. Two linear benchmarks are performed --- low $\beta$ tearing modes in W7-AS, as well as low and high toroidal mode number, $n$, interchange modes in a finite $\beta$, net zero current carrying stellarator geometry. The first test is based on experimental observations of such tearing modes in W7-AS discharges with ohmic current drive \cite{weller2001}, and used to demonstrate that the JOREK model can handle shaped stellarator geometries. The second is intended to investigate whether reduced MHD can be applied in the low magnetic shear regime. For the considered test case, while the acoustic wave and the influence of parallel flows needs to be included in order to capture the compressible dynamics correctly, particularly for low $n$ modes, compressional Alfv\'enic activity does not appear to play an important role. This implies that there is a region of validity, where the application of reduced MHD models without the influence of compressional Afv\'enic activity, can be justified. These benchmarks are carried out against the CASTOR3D code, a full MHD model including viscoresistive and extended MHD effects \cite{Strumberger2016, Strumberger2019}.

In Section \ref{sec:validation}, an initial experimental validation study is conducted using equilibrium reconstructions of W7-AS high $\beta$ discharges \cite{weller2003investigation}. Three experimental validation exercises can be identified with the given equilibria. Firstly, the MHD activity can be shown to be resistive, as characterised in the experiment. Secondly, the experimental evidence supports the conclusion that pressure driven MHD activity saturates at mild amplitudes, such that increased heating power can allow operation at $\beta$ values beyond the limit prescribed by Mercier's criterion and linear resistive interchange analysis. Thirdly, low $n$ resistive ballooning activity has been experimentally observed at intermediate $\beta$ values, which can be suppressed with increased heating power \cite{weller2003investigation}. This paper provides an initial attempt at reproducing these observations with JOREK.

Among the physical parameter scans that have been carried out, plasma resistivity is demonstrated to be particularly important in dictating whether the MHD activity leads to significant stochastisation over the plasma volume and a loss of confinement, or soft MHD activity that is unable to penetrate into the core of the device. In such a way, JOREK results corroborate the experimental observations that the governing MHD activity is viscoresistive. At parameters that approach experimental values, a soft $\beta$ limit appears to be observed. An initial observation of low $n$ MHD activity, particularly the (2, 1) resistive ballooning mode observed in the experiment, is also shown, as a starting point for future more detailed analysis. To conclude the paper, the strategy for further validation and interrogation of the underlying physical mechanism behind the simulated soft $\beta$ limit, the next steps for the development of the nonlinear model, and an outlook on future predictive $\beta$ limit studies are discussed in section \ref{sec:conclusion}. 

\section{The reduced MHD model for stellarators} \label{sec:reduced_mhd}
A detailed derivation of the original reduced MHD model implemented in JOREK is described in \cite{nikulsin2021models}. In this section, a brief overview of this model is provided, including the newly implemented parallel momentum equation, and modifications to the vacuum field representation. A two temperature model is also available, but is not applied in this study.

\subsection{Reduced MHD ansatz}
In the reduced MHD ansatz used herein, the magnetic field, $\mathbf{B}$, and velocity field, $\mathbf{v}$, are represented as

\begin{equation} \label{eq:magnetic_ansatz}
    \mathbf{B} = \nabla \chi + \nabla \Psi \times \nabla \chi
\end{equation}

\begin{equation} \label{eq:velocity_ansatz}
    \mathbf{v} = \frac{\nabla \Phi \times \nabla \chi}{B_\mathrm{v}^2} + \mathrm{v}_\parallel \mathbf{B},
\end{equation}  

where $\Psi$ is the magnetic stream function corresponding to the induced current along the vacuum magnetic field, $\nabla \chi$, and $B_\mathrm{v} = |\nabla \chi|$. 

Regarding equation \ref{eq:velocity_ansatz}, the potential function, $\Phi$, is used to represent the $\mathbf{E} \times \mathbf{B}$ velocity, and $\mathrm{v}_\parallel$ is the parallel flow velocity along the magnetic field lines. In the above equations, $\mathrm{v}_\parallel$ can be set to zero, recovering the original model used in \cite{nikulsin2022jorek3d}. This enables us to compare the validity of the model with and without the inclusion of the acoustic wave dynamics that are represented by this term.

\subsection{Initial equilibrium condition} \label{sec:init_equil_cond}
Before a particular simulation case can be evolved in time, the equilibrium condition must be formulated, using the ansatz in equation \ref{eq:magnetic_ansatz} and \ref{eq:velocity_ansatz}. Similar to \cite{nikulsin2022jorek3d}, detailed checks of the initial equilibrium reconstruction need to be carried out, to ensure that the starting condition for the initial value problem remains approximately valid as an equilibrium condition once it has been recomputed using the JOREK magnetic field ansatz. This is a challenge that all codes must contend with, when modifying the magnetic field representation used in ideal MHD codes such as VMEC and GVEC, such that the assumption of nested flux surfaces is relaxed. It is not guaranteed that the equilibrium force balance is preserved in the transformation to a different ansatz.

The steps to construct this initial condition are the same as described in the fourth paragraph of Section I in \cite{nikulsin2022jorek3d}, except for the calculation of the vacuum field and boundary condition for $\Psi$. Dommaschk potentials were originally used to represent $\chi$ \cite{dommaschk1986representations}. In the latest version of the code, it is also possible to use a finite element representation of the vacuum scalar potential. The vacuum field is calculated by solving a simple Poisson equation for $\chi$, enforcing that $\nabla \cdot \mathbf{B}_\mathrm{v}=0$, and that $\mathbf{n} \cdot \mathbf{B}_\mathrm{v}=0$ on the plasma boundary. With this modification, the correction to the free boundary vacuum field that enforces an ideal wall boundary condition is included in the vacuum field representation, $\nabla \chi$, itself. 

The finite element representation of $\chi$ has the advantage that it is easier to construct an accurate representation of the ideal wall boundary conditions for the total magnetic field in shaped geometries. The ideal wall boundary condition, $\mathbf{n} \cdot \mathbf{B}=0$, can be easily satisfied by setting $\Psi$ as a constant on the boundary. In such a way, the steps in Section IV of \cite{nikulsin2022jorek3d} can be circumvented. While the approach using Dommaschk potentials can in principle still be used to model optimised stellarators, numerical difficulties have been encountered when representing the vacuum field of highly shaped devices. The finite element representation is therefore used in all simulations which follow, as it is a more robust approach. 


\subsection{Time evolution equations} \label{sec:time_evol}
Once the equilibrium condition has been represented in the JOREK ansatz, the system needs to be evolved in time. In order to solve for the evolution of the field variables in equation \ref{eq:magnetic_ansatz} and \ref{eq:velocity_ansatz}, as well as the plasma density, $\rho$, and temperature, $T$, the reduced MHD ansatz is inserted into the single temperature viscoresistive MHD equations, resulting in the following reduced MHD equations:

\begin{eqnarray} \label{eq:density_equation}
    \eqalign{
    \frac{\partial \rho}{\partial t} = &-B_\mathrm{v} \left[ \frac{\rho}{B_\mathrm{v}^2}, \Phi \right] - B_\mathrm{v} \partial^\parallel (\rho \mathrm{v}_\parallel) \\ 
    & - B_\mathrm{v} \left[\rho \mathrm{v}_\parallel, \Psi \right] + P
    }
\end{eqnarray}

\begin{eqnarray}
    \eqalign{
    \nabla \cdot \left( \frac{\partial}{\partial t}\left[\frac{\rho}{B_\mathrm{v}^2}\nabla^\bot \Phi \right] \right) = &\frac{B_\mathrm{v}}{2} \left[ \frac{\rho}{B_\mathrm{v}^2}, \frac{\left(\Phi, \Phi\right)}{B_\mathrm{v}^2} \right] \\
                                & - B_\mathrm{v} \left[ \frac{\rho \omega}{B_\mathrm{v}^4}, \Phi \right] \\
                                & - \nabla \cdot \left[ \frac{\nabla^\bot \Phi}{B_\mathrm{v}^2}  \nabla \cdot \left(\rho \mathbf{v}\right) \right] \\
                                & + \nabla \cdot \left( j \mathbf{B} \right) + B_\mathrm{v} \left[ \frac{1}{B_\mathrm{v}^2}, p \right]\\
                                & + \nabla \cdot \left( \mu_\bot \nabla^\bot \omega \right) \\
                                &- \Delta ^\bot \left( \mu_\mathrm{num} \Delta^\bot \omega \right)
    }
\end{eqnarray}

\begin{eqnarray}
    \eqalign{
    \rho B^2 \frac{\partial \left(\rho \mathrm{v}_\parallel \right)}{\partial t} + \frac{\rho \mathrm{v}_\parallel}{2}\frac{\partial B^2}{\partial t} = & -\frac{\rho B_\mathrm{v}}{2}\partial^\parallel \mathrm{v}^2 - \frac{\rho B_\mathrm{v}}{2} \left[\mathrm{v}^2, \Psi \right] \\
                                & - \mathrm{v}_\parallel B^2 \nabla \cdot \left( \rho \mathbf{v} \right) \\
                                & - B_\mathrm{v} \partial^\parallel p - B_\mathrm{v} \left[p, \Psi\right]\\
                                & + \nabla \cdot \left[\mu_{\parallel, \bot} \nabla_\bot \mathrm{v}_\parallel + \mu_{\parallel, \parallel} \nabla_\parallel \mathrm{v}_\parallel\right]
    }
\end{eqnarray}

\begin{eqnarray}
    \eqalign{
        \frac{\partial \left(\rho T\right)}{\partial t} = &	-\frac{1}{B_\mathrm{v}}\left[ \rho T, \Phi \right] - \mathrm{v}_\parallel B_\mathrm{v} \partial^\parallel p - \mathrm{v}_\parallel B_\mathrm{v} \left[p, \Psi \right]\\
                                             &  - \Gamma \rho T B_\mathrm{v} \left[ \frac{1}{B_\mathrm{v}^2}, \Phi\right] \\
                                             &  - \Gamma p B_\mathrm{v} \partial^\parallel \mathrm{v}_\parallel - \Gamma p B_\mathrm{v} \left[ \mathrm{v}_\parallel, \Psi \right]\\                                   
                                             &  + \nabla \cdot \Bigg[ \Bigg. \kappa_\bot \nabla_\bot T + \kappa_\parallel  \nabla_\parallel T + T D_\perp \nabla_\bot \rho \\
                                             & +T D_\parallel \nabla_\parallel \rho  \Bigg.\Bigg] + \left( S_e + \eta_\mathrm{Ohm} (\Gamma - 1) B_\mathrm{v}^2 j^2\right)
    }
\end{eqnarray}

\begin{eqnarray} \label{eq:induction_equation}
    \eqalign{
        \frac{\partial \Psi}{\partial t} &= \frac{\partial^\parallel \Phi - \left[\Psi, \Phi\right]}{B_\mathrm{v}}  + \eta \left(j - j_\mathrm{source}\right) \\
                                         &- \nabla \cdot \left(\eta_\mathrm{num} \nabla^\bot j \right)
    }
\end{eqnarray}

where the auxiliary variables for the plasma current, $j$, and vorticity, $\omega$ are defined as

\begin{equation}
    j = \Delta^* \Psi
\end{equation}

\begin{equation}
    \omega = \Delta^\bot \Phi
\end{equation}

and $P = \nabla \cdot \left(D_\bot \nabla_\bot \rho + D_\parallel \nabla_\parallel \rho \right)+ S_\rho$. In the above equations, the operators are defined as follows

\begin{eqnarray*}
\begin{array}{cc}
\nabla_\parallel = B^{-2} \mathbf{B}(\mathbf{B}\cdot \nabla, & \nabla_\bot = \nabla - \nabla_\parallel \\
\partial^\parallel = B_\mathrm{v}^{-1} \nabla \chi \cdot \nabla, & \nabla^\bot = \nabla - B_\mathrm{v}^{-1} \nabla \chi \partial^\parallel \\
\Delta^\bot = \nabla \cdot \nabla^\bot, & \nabla^* = B_\mathrm{v}^{-2}\nabla \cdot (B_\mathrm{v}^2 \nabla^\bot \\
\left[f, g\right] = B_\mathrm{v}^{-1} \nabla \chi \cdot \left(\nabla f \times \nabla g \right), & (f, g) = \nabla^\bot f \cdot \nabla^\bot g
\end{array}    
\end{eqnarray*}

It should be noted that the components of the momentum equation above are written in conservative form, unlike in \cite{nikulsin2022jorek3d}. However, the equation was implemented in conservative form in the code for this previous study as well. Furthermore, the artificial parallel diffusivity, $D_\parallel$, is only used when the parallel flow is neglected, in order to approximate the redistribution of the plasma along magnetic field lines. 

The elimination of compressional Alfv\'enic activity in equations \ref{eq:density_equation} through \ref{eq:induction_equation} means that this reduced MHD model has a limited region of validity. For the tokamak reduced MHD model implemented in JOREK, the region of validity has been tested rigorously against full MHD models in previous work \cite{pamela2020extended}. Therein, it was shown that the finite $\beta$ (1, 1) internal kink mode is not captured by reduced MHD models, while most other instabilities of interest are captured correctly. 

A remaining concern that should be addressed by the stellarator model has been raised in \cite{graves2019reduced} regarding regions of negligible magnetic shear. In this analytical work, it is argued that the elimination of magnetic field or fluid compression in reduced MHD models means that the plasma dynamics can be prevented from minimising the compressional terms in the full MHD equations. In such a way, this leads to an artificial stabilisation of compressible modes, whether through unphysical fluid or magnetic field compression. 

This concern is particularly important for stellarators, which typically have rotational transform profiles with low magnetic shear, approaching or crossing low order rational surfaces. Low order rational surfaces are concerning because the modes associated with them will be broader. In such a way, the artificial stabilising effect of the compressional work performed by plasma instabilities will be larger, leading to more significant errors. This effect is assessed numerically in the results shown in Section \ref{sec:l2_interchange}.

\subsection{Numerical methods}
JOREK has a 2D, B\'ezier finite element representation of poloidal planes \cite{czarny2008bezier}, combined with a 1D spectral representation for both the grid and physics variables in the toroidal direction in stellarator applications. The finite elements are typically $G^1$ continuous, but the generalisation of the elements to higher order continuity in \cite{pamela2022generalised} could become important for the stellarator model, as it might follow to eliminate auxiliary variables, $j$ and $\omega$, from the system. This feature has not been used in the current work.

It should be noted that simulations can either be carried out over the full torus, or a single field period. Single field period simulations neglect symmetry breaking modes, such that they are much less computationally intensive, and thus attractive for preliminary studies. Full torus simulations are typically necessary to capture the full nonlinear dynamics, particularly of low $n$ modes.   

The equations are solved in weak form, and evolved in time using an implicit time stepping scheme. For stellarator simulations, the mode group preconditioning developed in \cite{holod2021enhanced} has been important for making simulations computationally manageable. By coupling adjacent toroidal harmonics in the same mode family together in the preconditioner, some of the linear toroidal mode coupling of toroidally shaped devices can be accommodated, enhancing convergence of the implicit solver. The mode coupling is nevertheless often strong in more optimised configurations, even in the second and third toroidal sidebands. As a result, further improvements are part of ongoing work to make stellarator simulations as computationally tractable as equivalent tokamak runs.


\section{Verification studies} \label{sec:verification}

The following section aims to justify the use of the reduced MHD model outlined in Section \ref{sec:reduced_mhd} for studying strongly shaped geometries, and finite $\beta$ instabilities in regions of low magnetic shear. Two test cases are constructed to consider these regimes separately - a (2, 1) tearing mode in W7-AS geometry, and interchange modes in a net zero current carrying stellarator.

\subsection{(2, 1) tearing mode in W7-AS geometry} \label{sec:w7as_tearing}

\begin{figure}
    \includegraphics[width=0.5\textwidth]{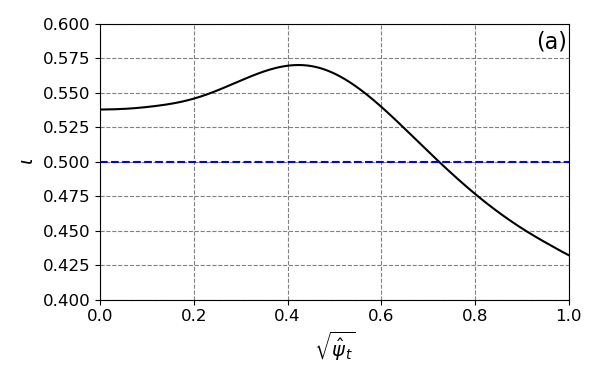}
    \includegraphics[width=0.5\textwidth]{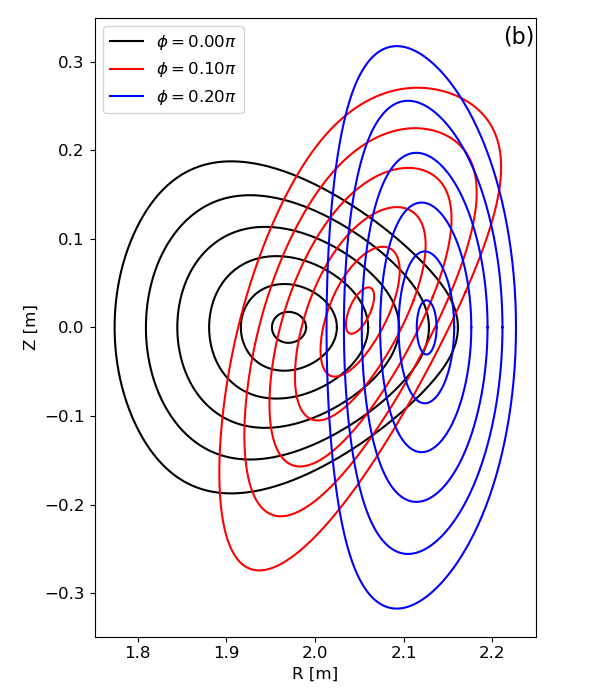}
    \caption{Rotational transform profile (a) and initial equilibrium flux surfaces (b) for W7-AS tearing mode test case.}
    \label{fig:w7as_tearing_equilibrium}
\end{figure}

The first test case is inspired by past experimental observations of tearing modes in W7-AS, when Ohmic plasma currents are driven in the device \cite{weller2001}. The rotational transform profile, and initial equilibrium flux surfaces of the test case are shown in Figure \ref{fig:w7as_tearing_equilibrium}. Rather than using an exact equilibrium reconstruction informed by experimental data, the test case is an approximate reconstruction generated by adding the toroidal current profile from Figure 28 of \cite{weller2001} to an appropriate W7-AS equilibrium operating with external rotational transform, $\iota_\mathrm{ext}\approx0.34$. The additional toroidal current increases $\iota$ such that there is a $\iota=0.5$ rational surface within the plasma volume, making it unstable to a (2, 1) tearing mode.

The baseline simulations are run with the parameters shown in Table \ref{tab:baseline_tearing}. It should be noted that for this low $\beta$ current driven mode, $\mathrm{v}_\parallel$ is not necessary to capture the instability, so equation (5) is excluded from the model and we set $\mathrm{v}_\parallel = 0$ in the other equations. For the linear benchmark with CASTOR3D, the simulations are carried out in the no viscosity limit. To get a physical scan of resistivity in this limit, the parameter, $k$, is used to vary the viscoresistive parameters together, keeping the magnetic Prandtl number constant across simulations. Isotropic, relatively low diffusivity parameters are prescribed to ensure that the equilibrium does not change over the linear phase.

\begin{table}
\centering
\caption{\footnotesize{Physical and resolution parameters used in the linear (2, 1) tearing mode benchmark between JOREK and CASTOR3D shown in Figure \ref{fig:w7as_linear_growth_rate}. Nonlinear parameters are also shown for the results shown in Figure \ref{fig:w7as_tearing_nonlin}. All parameters are constant across the simulated volume. For the resistivity scan, $k$ is used to vary viscoresistive parameters together in the no viscosity limit.}}
\begin{tabular}{c|c|c}
Parameter                                     & Linear Benchmark                  &   Nonlinear run   \\ \hline
      $\mathrm{n}$                                     & $10^{20}$                         &   $10^{20}$              \\
      $\eta\ [\Omega \mathrm{m}]$                      & $k \times 1.93 \times 10^{-6}$    &   $1.93 \times 10^{-6}$  \\
      $\eta_{\mathrm{num}}\ [\Omega \mathrm{m}^3]$     & $k \times 1.93 \times 10^{-12}$   &   $1.93 \times 10^{-12}$ \\
      $\mu_\perp\ [\mathrm{kgm}^{-1}\mathrm{s}^{-1}]$                 & $k \times 5.15 \times 10^{-10}$   &   $5.15 \times 10^{-8}$  \\
      $\mu_{\mathrm{num}}\ [\mathrm{kgms}^{-1}]$       & $k \times 5.15 \times 10^{-16}$   &   $5.15 \times 10^{-14}$ \\
      $D_\bot [\mathrm{m}^2\mathrm{s}^{-1}]$                    & $0.154$                           &   $0.154$  \\
      $D_\parallel [\mathrm{m}^2\mathrm{s}^{-1}]$               & $0.154$                           &   $0.154$  \\
      $\kappa_\bot [\mathrm{m}^2\mathrm{s}^{-1}]$               & $0.231$                           &   $0.231$  \\
      $\kappa_\parallel [\mathrm{m}^2\mathrm{s}^{-1}]$          & $0.231$                           &   $0.231$  \\
      $\mathrm{n}_\mathrm{rad}$                               & $64$                                &    64        \\
      $\mathrm{n}_\mathrm{pol}$                               & $64$                                &    64        \\
      $\mathrm{n}_\mathrm{tor}$                               & 0,1,...39,40                              &    0,1,...,19,20        \\ 
\end{tabular}
\label{tab:baseline_tearing}
\end{table}

\begin{figure}
    \centering
    \includegraphics[width=0.5\textwidth]{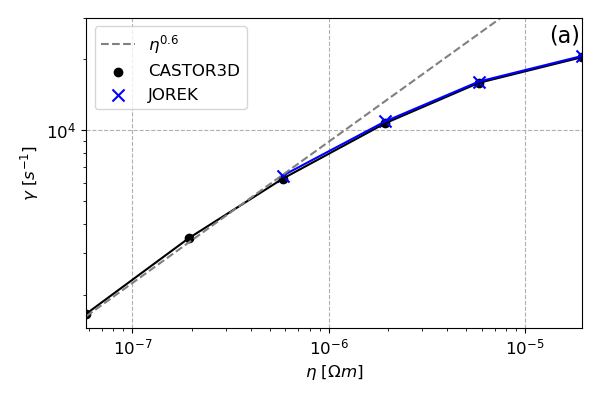}
    \includegraphics[width=0.5\textwidth]{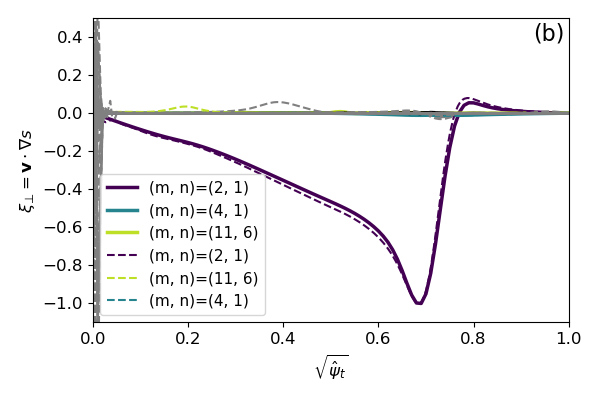}
    \caption{Linear growth rate comparison of (2, 1) resistive tearing modes in W7-AS (a). CASTOR3D results are continued to low resistivity values, where the typical scaling for resistive tearing modes is recovered. The linear eigenfunctions computed in JOREK (solid) and CASTOR3D (dashed) using Boozer coordinates for the $\eta=1.93\times 10^{-6}\ \Omega \mathrm{m}$ case (b) also agree well.}
    \label{fig:w7as_linear_growth_rate}
\end{figure}

The linear growth rate and eigenfunction comparison is shown in Figure \ref{fig:w7as_linear_growth_rate}. The agreement between the reduced and full MHD approaches is very good, with the maximum deviation in the growth rates being approximately $3\ \%$. JOREK results are not carried out at lower resistivities because lower values can be numerically challenging while running in the no viscosity limit, and require longer timescales to resolve the linear phase. Note that the match in growth rates is actually better than in the simpler test case, studied previously in \cite{nikulsin2022jorek3d}, where a mismatch is found at higher resistivity. This is because it was found that the helical mode coupling was not correctly resolved in the initial study, due to the use of insufficient poloidal resolution in CASTOR3D. At higher poloidal resolution, the level of agreement for this simpler case is similar to the results shown herein.

The linear eigenfunction in Figure \ref{fig:w7as_linear_growth_rate} (b) is computed in Boozer coordinates for the $\eta=1.93\times10^{-6}\ \Omega\mathrm{m}$ case. The dominant (2, 1) Fourier mode shows the typical tearing structure, and is in very good agreement between the two codes. The poloidal and toroidal sidebands in the CASTOR3D eigenfunction are marginally larger, but are below 6\%, and so this discrepancy is not considered significant. 

The simulation is continued into the nonlinear phase for the $\eta=1.93\times10^{-5}\ \Omega\mathrm{m}$ case to demonstrate that the nonlinear dynamics can be simulated without numerical problems. This resistivity is chosen because the timescale of the instability is shorter, such that the simulation is less computationally expensive.  As shown in Table \ref{tab:baseline_tearing}, the viscous parameters are increased, and the toroidal resolution, $\mathrm{n}_\mathrm{tor}$, referring to the highest toroidal mode number in the simulation, is decreased to 20. This is done to reduce the computational cost, and ensure numerical stability in the nonlinear phase.

\begin{figure}
    \centering
    \includegraphics[width=0.5\textwidth]{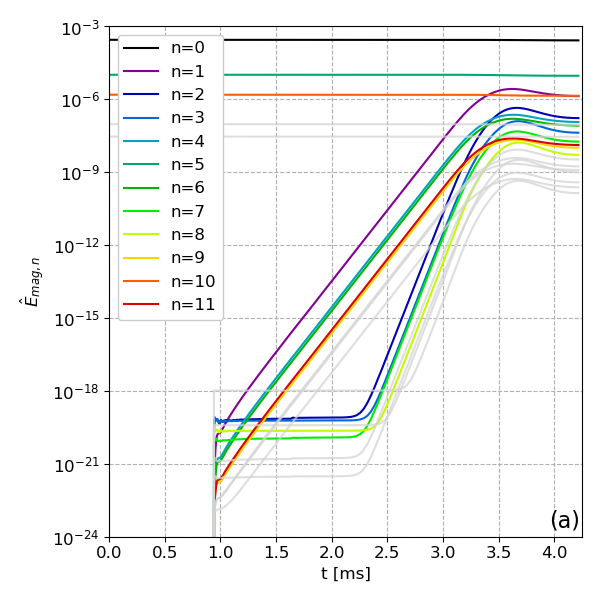}
    \includegraphics[width=0.5\textwidth]{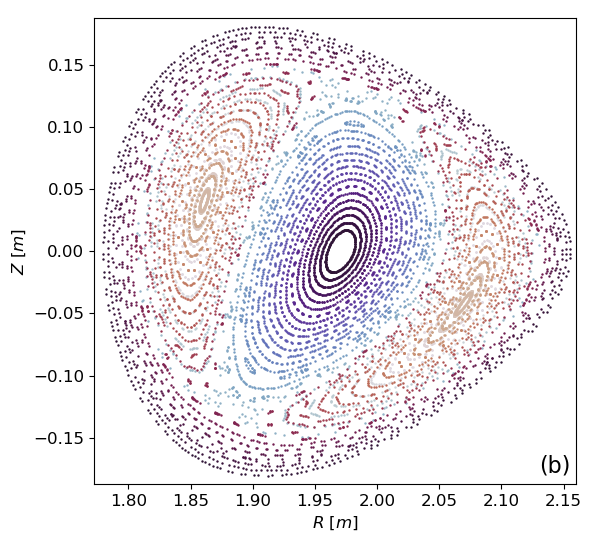}
    \caption{Normalised magnetic energy traces (a) of (2, 1) tearing mode in W7-AS with $\eta=1.93\times10^{-5}\ \Omega \mathrm{m}$ using the nonlinear parameters in Table \ref{tab:baseline_tearing}. Higher sub-dominant toroidal modes are coloured in light gray. A Poincar\'e plot (b) shows the saturated (2, 1) island structure at $t=4.18\ \mathrm{ms}$.}
    \label{fig:w7as_tearing_nonlin}
\end{figure}

The results of this nonlinear run are shown in Figure \ref{fig:w7as_tearing_nonlin}. Note that herein, when magnetic energy traces of individual toroidal harmonics are shown, the calculated energies are approximated by the following integrals 

\begin{equation}
    \hat E_{mag,n} = \int B_\mathrm{v}^2(\Psi_n, \Psi_n) J(s,t,\phi) ds dt d\phi,
\end{equation}

\begin{equation}
    \hat E_{kin,n} = \int \frac{(\Phi_n, \Phi_n)}{B_\mathrm{v}^2} J(s,t,\phi) ds dt d\phi, 
\end{equation}

where $J(s,t,\phi)$ is the Jacobian from local finite element coordinates to real space, and $s$, $t$ and $\phi$ are the radial, poloidal and toroidal basis vectors, respectively. This calculation neglects the cross terms produced by the toroidal variation of the Jacobian, but is sufficient to show the main features of the nonlinear saturation. When the total magnetic energy, $\hat E_{mag}$,  in a given mode family is calculated in Section \ref{sec:validation}, the cross terms are taken into account. 

The magnetic energy traces in Figure \ref{fig:w7as_tearing_nonlin} (a) show that the full torus calculation is initialised after an initial single field period run, in order to relax the initial force imbalance introduced by the equilibrium import from GVEC \cite{Hindenlang2019}. Given that W7-AS is a five field period stellarator, the toroidal harmonics can be separated into three separate toroidal mode families \cite{schwab1993ideal}, $N_\mathrm{f}=0$, corresponding to the equilibrium harmonics, and $N_\mathrm{f}=1$ and 2, corresponding to separate families of symmetry breaking modes. The magnetic energy traces show that the $N_\mathrm{f}=1$ mode family leads the perturbation. The adjacent toroidal sidebands in the $N_\mathrm{f}=1$ mode family, namely the $n=4$ and $n=6$ modes are the second largest mode contributions, followed by the $n=9$ and $n=11$ modes as one would expect. Much like a simple tearing mode in a tokamak, the $N_\mathrm{f}=2$ mode family is nonlinearly driven as the $N_\mathrm{f}=1$ modes saturate. 

At $t=4.18\ \mathrm{ms}$ near the end of the simulation time, the expected (2, 1) magnetic island structure can be seen in Figure \ref{fig:w7as_tearing_nonlin}. Note that in this Poincar\'e plot and all others shown herein, each magnetic field line is given a distinct colour in order to more easily distinguish the magnetic field structure. Given the above results, this benchmark is considered to have been successful in demonstrating the capability of JOREK to capture low $n$ current driven modes in stellarators with stronger shaping than previous studies.

\subsection{Interchange modes in a low magnetic shear stellarator} \label{sec:l2_interchange}

\begin{figure}
    \centering
    \includegraphics[width=0.5\textwidth]{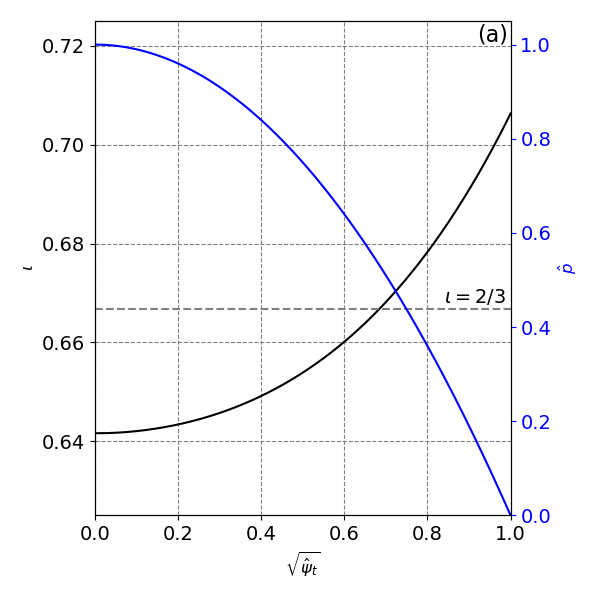}
    \includegraphics[width=0.5\textwidth]{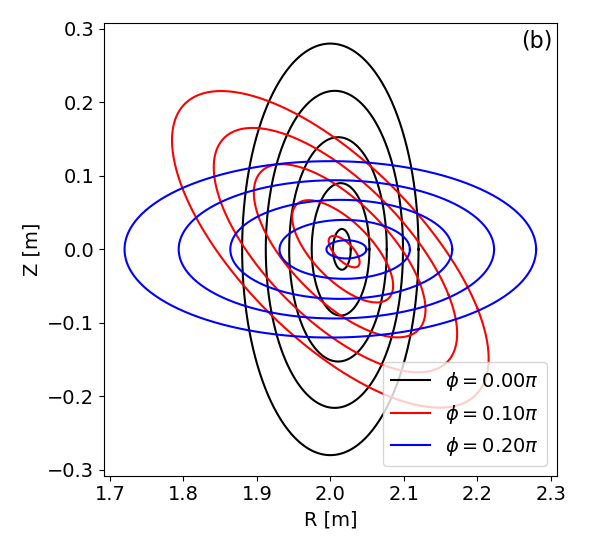}
    \caption{Equilibrium profiles (a) of the rotational transform (black) and pressure (blue), and equilibrium flux surfaces (b) of $l=2$ finite $\beta$ interchange case.}
    \label{fig:l2_interchange_equilibrium}
\end{figure}

To test the low magnetic shear regime, a finite $\beta$, five field period, $l=2$ classical stellarator equilibrium was constructed, which is unstable to low and high $n$ interchange modes. This case is a modified version of the equilibrium reconstruction of the W7-A stellarator used in \cite{nikulsin2022jorek3d}. The rotational transform, pressure profile and equilibrium flux surfaces are shown in Figure \ref{fig:l2_interchange_equilibrium}. In particular, the rotational transform crosses the $\iota=2/3$ low order rational surface making it highly unstable. While the test case is a classical stellarator, to get to this value of rotational transform without net toroidal current, strong elliptical shaping is required, such that the ratio of the major to minor radius of its elliptical cross section is 1.6. This level of strong toroidal shaping was not achieved experimentally in W7-A, which had a maximum external rotational transform, $\iota_\mathrm{ext} \approx 0.21$ \cite{team1980}.

\begin{figure}
    \centering
    \includegraphics[width=0.5\textwidth]{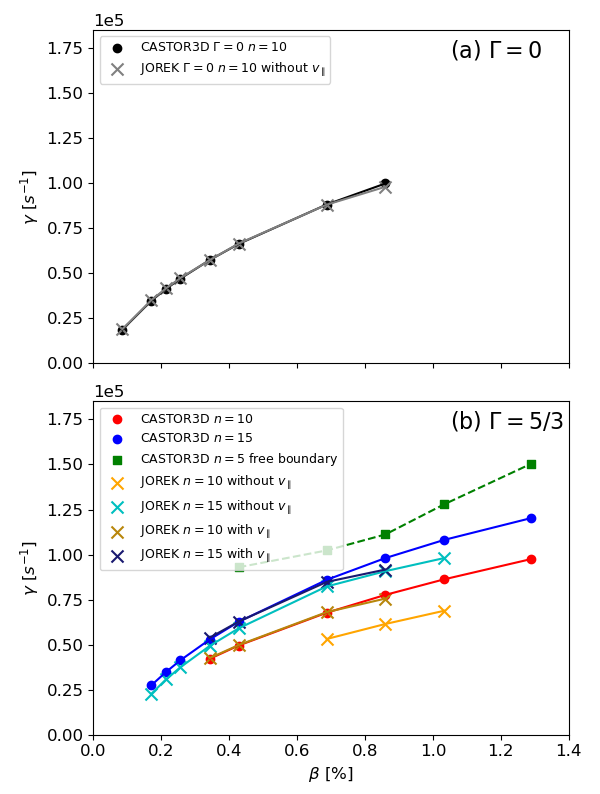}
    \caption{Linear benchmark of (15, 10) and (22, 15) interchange modes excluding (a) and including (b) the influence of compressibility by varying $\Gamma$. A comparison of growth rates in the compressible regime, neglecting parallel flow terms in JOREK is also shown in (b) in order to show the stabilising effect from artificial fluid compression.}
    \label{fig:l2_beta_benchmark}
\end{figure}

To demonstrate the importance of the parallel flow terms in modeling compressible effects, a $\beta$ scan is conducted by linearly scaling the pressure profile, keeping the $\iota$ profile constant, as shown in Figure \ref{fig:l2_beta_benchmark}. Growth rates are once again compared with CASTOR3D. The simulation parameters are shown in Table \ref{tab:l2_parameters} and approach the ideal limit, as the instability is an ideal mode \cite{nuehrenberg2023}. As many simulations need to be run for the $\beta$ scan, it has been restricted to single field period calculations, which are more computationally tractable. In the incompressible limit, shown in Figure \ref{fig:l2_beta_benchmark} (a), the growth rates of the reduced MHD model without parallel flows agrees very well with full MHD results. Furthermore, the point of marginal stability is recovered. In Figure \ref{fig:l2_beta_benchmark} (b), the artificial stabilisation error that comes from neglecting parallel flows in compressible modes is shown. While the expected decrease in the linear growth rate of the $n=10$ and $n=15$ mode is observed when $\Gamma=5/3$, the stabilisation in JOREK is larger and does not agree with CASTOR3D. This is the artificial stabilisation effect discussed in Section \ref{sec:time_evol}. 

\begin{table}[]
    \centering
    \caption{Physical and resolution parameters for the linear benchmark with CASTOR3D in Figure \ref{fig:l2_beta_benchmark} and \ref{fig:l2_n_benchmark}, and the nonlinear simulation shown in Figure \ref{fig:l2_75_external}. For the nonlinear run, the resistivity and viscosity profiles have a Spitzer-like temperature dependence, and the density profile is proportional to the square root of the pressure profile to obtain a reasonable vacuum region around the plasma. All other parameters are constant across the plasma volume, including the initial density in the linear benchmark case.}
    \begin{tabular}{c|c|c}
      Parameter & Linear benchmark & Nonlinear run  \\ \hline
              $n$                                                  & $10^{20}$              &  $10^{20}$              \\
              $\eta\ [\Omega \mathrm{m}]$                                   & $1.93 \times 10^{-8}$  &  $1.93 \times 10^{-8}$  \\
              $\eta_{\mathrm{num}}\ [\Omega \mathrm{m}^3]$                  & 0                      &  $1.93 \times 10^{-14}$ \\
              $\mu_\perp\ [\mathrm{kgm}^{-1}\mathrm{s}^{-1}]$                              & 0                      &  $5.15 \times 10^{-11}$ \\
              $\mu_{\mathrm{num}}\ [\mathrm{kgms}^{-1}]$                    & 0                      &  $5.15 \times 10^{-17}$ \\
              $\mu_{\parallel, \bot}\ [\mathrm{kgm}^{-1}\mathrm{s}^{-1}]$                    & 0                      &  $5.15 \times 10^{-8}$  \\
              $\mu_{\parallel, \parallel}\ [\mathrm{kgm}^{-1}\mathrm{s}^{-1}]$             & 0                      &  $5.15 \times 10^{-8}$  \\
              $D_\bot [\mathrm{m}^2\mathrm{s}^{-1}]$                                 & $0.0154$               &  $1.54$                 \\
              $\kappa_\bot [\mathrm{m}^2\mathrm{s}^{-1}]$                            & $0.0231$               &  $2.31$                 \\
              $\kappa_\parallel [\mathrm{m}^2\mathrm{s}^{-1}]$                       & $0.0231$               &  $2310000$              \\
              $\mathrm{n}_\mathrm{rad}$                                            & $71$                   &    $71$                 \\
              $\mathrm{n}_\mathrm{pol}$                                            & $96$                   &    $96$                 \\
              $\mathrm{n}_\mathrm{tor}$                                            & 0,5,...25,30                   &    0,5,...25,30                    
      \end{tabular}
      \label{tab:l2_parameters}
\end{table}

When parallel flows are included, good agreement is recovered for the $n=10$ and $n=15$ modes. Note that these distinct modes can both be seen in the same simulation run. It is clearly seen that there is a slightly larger finite error for the $\beta=0.86\ \%$ case, which is the largest $\beta$ value simulated in JOREK. The reason for this has been identified as a (7, 5) numerical instability close to the boundary of the simulation domain, which interferes with the $n=10$ and $n=15$ modes. The (7, 5) structure is sub-dominant in the $\beta=0.86\ \%$ case, such that it has only been observed in the Fourier decomposition of $j$ in the nonlinear phase (not shown). At larger $\beta$, the numerical structure grows and distorts the background field, before the internal modes can develop, resulting in a lower growth rate.

\begin{figure*}[ht]
    \centering
    \includegraphics[width=\textwidth]{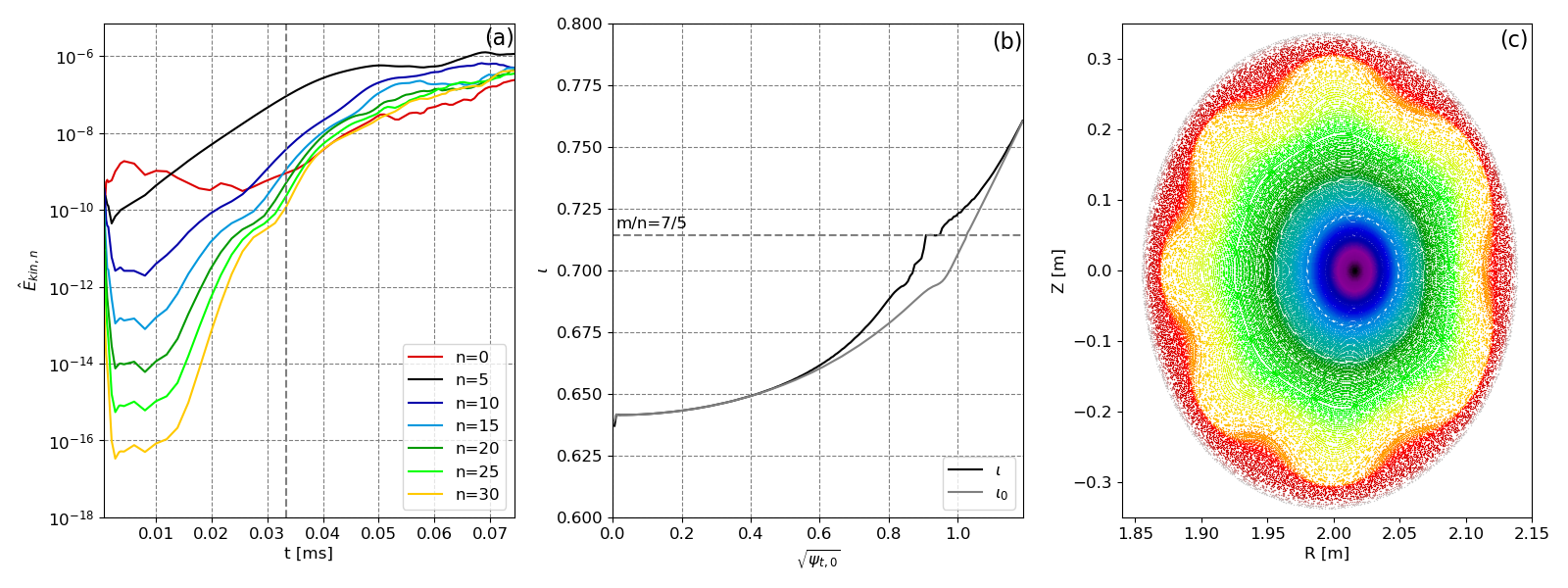}
    \caption{Normalised kinetic energy traces (a), $\iota$ profile diagnostic calculated using field line tracing (b) and Poincar\'e plot (c) of the initial saturation of the (7, 5) mode, marked by a grey dashed line in (a). The radial coordinate of the $\iota$ profile in (b) uses the initial toroidal flux, $\psi_{\mathrm{t},0}$. This explains why the (7, 5) resonance is shifted inwards, along with the displacement of the flux surfaces on the low field side midplane. Note that compared to Figure \ref{fig:l2_interchange_equilibrium} (b), the aspect ratio is not preserved in (c). This is done in order to show the magnetic field structure more clearly.}
    \label{fig:l2_75_external}
\end{figure*}

This numerical structure corresponds to a physical external instability, which has been identified in CASTOR3D, as shown in Figure \ref{fig:l2_beta_benchmark} (b). Numerical structures can appear in JOREK when such strong low mode number instabilities are externally unstable, such that the benchmark for the internal modes could not be carried out at higher $\beta$. By extending the VMEC equilibrium into the vacuum region, in a similar way to previous work on external kinks \cite{ramasamy2023eps}, the (7, 5) external mode can also be observed in JOREK, as shown in Figure \ref{fig:l2_75_external}. The simulation parameters for this nonlinear run are shown in table \ref{tab:l2_parameters}. Note that the resistivity has a Spitzer-like dependence for this simulation, ensuring that the low temperature vacuum region is resistive and does not significantly damp the mode growth.

The energies in Figure \ref{fig:l2_75_external} (a) show that the (7, 5) mode grows within 10s of microseconds. While this growth rate is comparable to that from CASTOR3D in Figure \ref{fig:l2_beta_benchmark}, a rigorous linear benchmark of this mode has not been attempted, as the treatment of the external region and boundary conditions in JOREK and CASTOR3D are very different. A free boundary implementation would be necessary in JOREK before this can be attempted, like the ones available for tokamak applications. 

The change to the $\iota$ profile is assessed during the initial saturation using field line tracing. Compared to the initial external rotational transform, $\iota_0$, the 7/5 resonance is shifted inwards. This is because the radial coordinate is computed from the initial toroidal flux, $\psi_{\mathrm{t},0}$, taking points along the midplane, extending from the magnetic axis to the simulation boundary on the low field side. It can be seen in Figure \ref{fig:l2_75_external} (c) that the flux surfaces are radially perturbed inwards at the outboard midplane, so that the external 7/5 resonance moves into the initial plasma region. A small external (7, 5) magnetic island structure can be observed in JOREK, corresponding to a flattening of the $\iota$ profile around the 7/5 resonance. This external island structure grows larger later in time, with strong MHD activity leading to a loss in confinement across most of the plasma volume (not shown).

\begin{figure}
    \centering
    \includegraphics[width=0.485\textwidth]{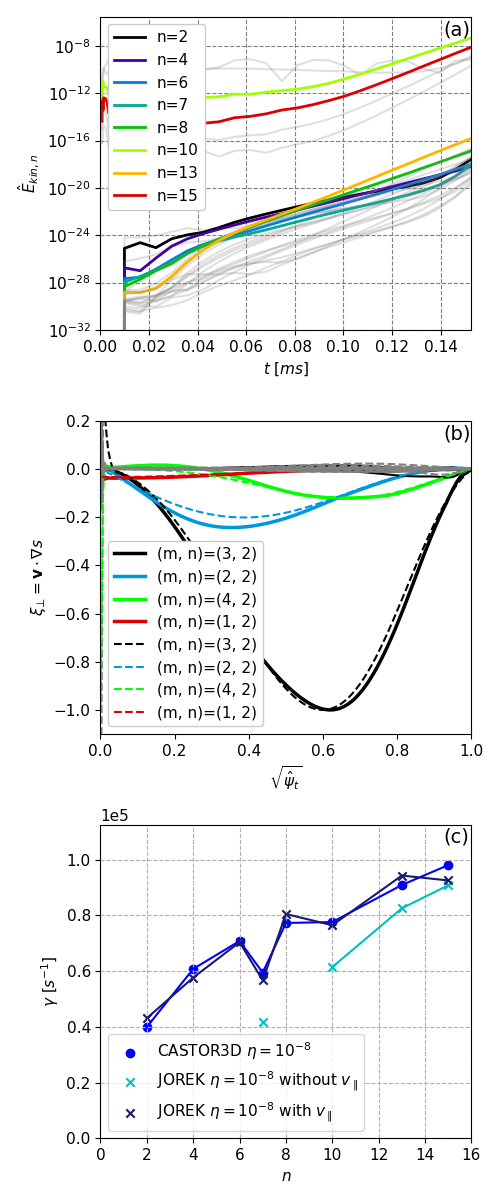}
    \caption{Linear benchmark of interchange modes for $\beta=0.86\ \%$ case. Multiple independent instabilities grow early in the full torus JOREK simulation, including parallel flows (a). Higher sub-dominant toroidal modes are coloured in light gray. The dominant $n=2$ Fourier harmonics have the same radial structure as the (3, 2) linear instability found in CASTOR3D (b) during this phase, taken at $t\approx0.1\ \mathrm{ms}$. The extracted growth rate of all modes identified in JOREK agree within $10\%$, when parallel flows are included (c).}
    \label{fig:l2_n_benchmark}
\end{figure}

To further support that compressional Alfv\'enic activity can be neglected in the simulated case, a further linear toroidal mode number scan of all observable modes in full torus JOREK runs for the $\beta=0.86\ \%$ case is carried out. The results are shown in Figure \ref{fig:l2_n_benchmark}. First, considering the evolution of the spectrally decomposed kinetic energies in Figure \ref{fig:l2_n_benchmark} (a), many toroidal harmonics grow at different rates in the early linear phase, prior to $t \approx 0.14\ \mathrm{ms}$. This indicates that multiple linearly unstable modes can be identified within the same simulation. For example, to demonstrate that the growth of the $n=2$ mode in JOREK corresponds to its own linear instability, the mode structure at $t=0.1\ \mathrm{ms}$ is shown in Figure \ref{fig:l2_n_benchmark} (b), comparing with the linear (3, 2) mode found in CASTOR3D. There is good agreement in the four largest (m, n) contributions to the instability, supporting the conclusion that the growth rate in JOREK corresponds to this mode.   

The comparison of the growth rates in the two codes is shown in Figure \ref{fig:l2_n_benchmark} (c). When the parallel flows are included, the error in the growth rate of the modes varies below $10\ \%$, indicating good agreement. The results without the parallel flow dynamics are again included to show how the artificial stabilisation of pressure driven instabilities is stronger for low $n$ modes. In the simulation without parallel flows, the low $n$ interchange modes, particularly the (3, 2) mode, were not observed. The error in the observed growth rates is shown to increase as the mode number is reduced in Figure \ref{fig:l2_n_benchmark} (c). The intuitive reason for this is that the low $n$ modes are radially broader and more global, leading to a greater artificial compressional effect when parallel flow dynamics are neglected. Similar to the axial compression of a tube of fluid, parallel flows are necessary to minimise the fluid compression, allowing the instability to grow.

It is worth noting that the error does not increase with decreasing mode number in the case with $\mathrm{v}_\parallel$, and the growth rate of the (3, 2) mode observed in JOREK is actually slightly larger than the full MHD results. This is important because if the compressional Alfv\'en wave contributed to the dynamics of the test case, a similar, albeit less significant trend to what is observed in the case without $\mathrm{v}_\parallel$ would be expected. As this is not the case, it seems that the eliminated flux compression terms are not playing an important role in the given test case, and the reduced model in JOREK is capable of simulating these modes. 


\section{Validation of $\beta$ limits in W7-AS} \label{sec:validation}
The analysis in Section \ref{sec:verification} demonstrates the capabilities of JOREK in handling finite $\beta$ effects. A natural next step is to pursue validation studies, reproducing experimental observations of pressure driven MHD activity. W7-AS $\beta$ scans are targeted as a first step. W7-AS is particularly attractive for initial validation purposes. As discussed in \cite{hirsch2008major}, high $\beta$ operation was carried out at low magnetic field strength, $\approx 0.9\ \mathrm{T}$, and at relatively high densities $~2\times 10^{20}\ \mathrm{m}^{-3}$, such that the temperature of such discharges is low, in the order of $100\ \mathrm{eV}$. This means that the plasma parameters are less computationally demanding, compared to high $\beta$, high performance scenarios where extended MHD effects may also need to be taken into account.

Herein, the soft $\beta$ limit observed in \cite{weller2003investigation} is targeted. Mild low $n$ MHD activity was observed in the experiment when operating at intermediate $\beta$ \cite{zarnstorff2005equilibrium}. Linear computational analysis showed that the MHD equilibria were resistively unstable. The observed MHD activity could be overcome with increased heating power, eventually being suppressed when $\beta$ exceeded approximately 2.5 \%. In such a way, the maximum $\beta$ achieved was prescribed by enhanced transport rather than disruptive MHD effects.

Based on these experimental observations, there are three ways to demonstrate that JOREK can correctly capture the physics observed in the experiment:

\renewcommand{\theenumi}{\roman{enumi}}
\begin{enumerate}
  \setlength{\itemsep}{0pt}
  \item confirm that the observed MHD activity is resistive.
  \item using a heat source corresponding to the experimental heating power, show that the $\beta$ limit is not hard, or prescribed by MHD crash dynamics. 
  \item observe nonlinearly dominant low $n$ mode activity at $1.5\% < \beta < 2.5\%$.
\end{enumerate}

Regarding (iii), a (2, 1) resistive ballooning mode was observed in the experiment, and seen to saturate at mild values \cite{weller2003investigation}, which is treated as the expected result for full torus simulations of the case studied in this section. 

In this initial study, we focus on results towards these three validation goals. Single field period calculations are used to perform larger parameter scans at reasonable computational expense towards the first two goals. Using single field period calculations to study the $\beta$ limit can be justified in these studies, as it is known from the experiment that the low $n$ modes should not play a significant role in the high $\beta$ dynamics. In such a way, the single field period calculations are a necessary but not sufficient condition to demonstrate that JOREK captures the correct physical behaviour. To interrogate the third goal of the study, full torus simulations are necessary.

In Section \ref{sec:w7as_beta_equilib}, the ideal MHD equilibria used as an initial condition for this study are outlined. The MHD activity observed in JOREK is shown to be resistive in Section \ref{sec:w7as_beta_scan}, by carrying out a simplified $\beta$ scan, where the physical coefficients are treated as constant over the domain and independent of the plasma state. A soft $\beta$ limit is observed when the resistivity is sufficiently low, corresponding to weaker stochastisation of the plasma. Scans using physical parameters that approximate experimentally relevant values are then used in Section \ref{sec:w7as_experimentally_relevant} to support the conclusion that a soft $\beta$ limit should be observed under realistic conditions. An initial attempt at observing low $n$ MHD activity at intermediate $\beta$ values is shown in Section \ref{sec:w7as_low_n_activity}. Several of the limitations of the current modeling and directions for future work are discussed in Section \ref{sec:w7as_discussion}.

\subsection{Equilibrium reconstructions} \label{sec:w7as_beta_equilib}
\begin{figure}
    \centering
    \includegraphics[width=0.5\textwidth]{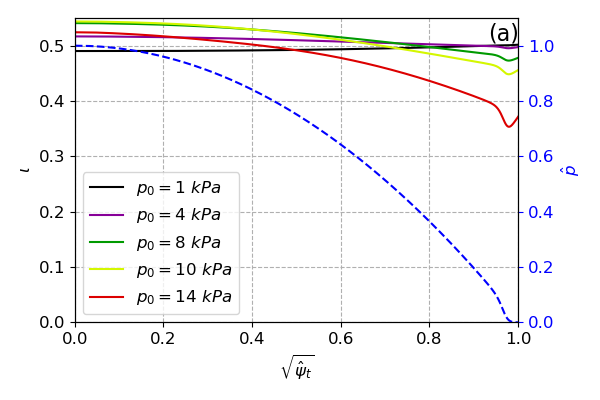}
    \includegraphics[width=0.5\textwidth]{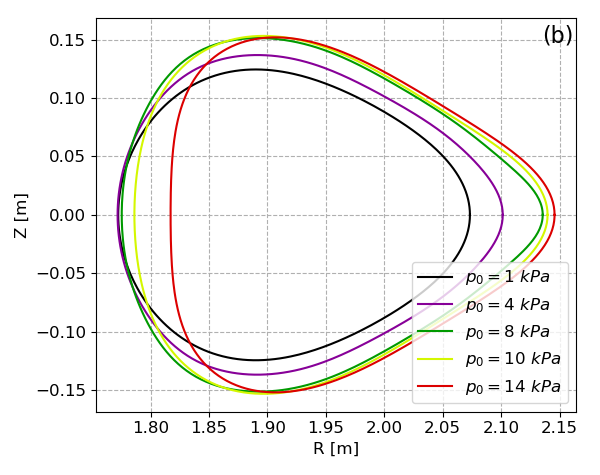}
    \caption{Normalised pressure profile (blue, dashed) and $\iota$ profiles (a) of W7-AS $\beta$ scan. The $\iota$ profiles show the development of tokamak-like shear with increasing $\beta$, and the last closed flux surfaces in the $\phi=0$ plane (b) exhibit the expected Shafranov shift.}
    \label{fig:w7as_beta_prof}
\end{figure}

The equilibria used in this study are free boundary VMEC equilibrium reconstructions of shot \#51751 on W7-AS \cite{weller2003investigation}, at different plasma $\beta$. The equilibrium profiles and last closed flux surface in the $\phi=0$ plane are shown in Figure \ref{fig:w7as_beta_prof} for different cases. The pressure profile is linearly scaled to modify $\beta$ in this scan. It can be seen that tokamak-like shear and a noticeable Shafranov shift develop with increasing pressure, just as is experimentally observed. The origin of these changes is the increase in Pfirsch-Schl\"uter currents in this partially optimised stellarator configuration.

The challenge of reconstructing the equilibrium using the JOREK magnetic field ansatz was described in Section \ref{sec:init_equil_cond}. For the equilibria shown in Figure \ref{fig:w7as_beta_prof}, numerical difficulties were found in reconstructing the $14\ \mathrm{kPa}$ case in the JOREK ansatz with sufficient accuracy that solving the initial value problem could be justified. A topic of further work is to improve the equilibrium reconstruction in JOREK, such that cases with such a strong Shafranov shift can be reconstructed accurately. For this reason, the analysis herein focuses on the cases with $\le 10\ \mathrm{kPa}$, where the equilibrium force balance and initial nested flux surfaces were reasonably preserved.

\subsection{Simplified $\beta$ scan} \label{sec:w7as_beta_scan}
Starting from the initial equilibria in Section \ref{sec:w7as_beta_equilib}. JOREK is used to evolve the simulations using the baseline parameters shown in the central column of Table \ref{tab:w7as_beta}. For this simplified analysis, the viscoresistive and diffusive parameters are kept constant and independent of the plasma state, and the initial density is $10^{20}\ \mathrm{m}^{-3}$ across the plasma volume. Anisotropic diffusion is used such that any ergodisation of the plasma should enhance the radial transport. A uniform heat source is prescribed with comparable heating power to the NBI heating of the experiment, $3.2\ \mathrm{MW}$, such that the plasma $\beta$ is expected to increase. A parallel momentum source is not necessary, as two counter-directional NBI sources were used for these high $\beta$ discharges.

It should be noted that the simulations are carried out without parallel flows. While the neglect of parallel flows leads to some stabilisation of the MHD activity, as shown in Section \ref{sec:l2_interchange}, the initial scan is only intended to demonstrate the main characteristics of the simulated case, and in particular, that the MHD activity is resistive. More experimentally relevant parameters are used in Section \ref{sec:w7as_experimentally_relevant}, including the influence of parallel flows.  

\begin{table}
    \centering
    \caption{Physical and resolution parameters for baseline W7-AS $\beta$ scan (center) in Section \ref{sec:w7as_beta_scan}, and approximate experimental values (right) for the $10\ \mathrm{kPa}$ case used in Section \ref{sec:w7as_experimentally_relevant}. The approximate experimental values are computed at the magnetic axis of the initial equilibrium.}
    \begin{tabular}{c|c|c}
    Parameter & Simplified $\beta$ scan & Exp. Approx.  \\ \hline
            $T\ [\mathrm{keV}]$                              & 0.061-0.864            &    0.31                   \\
            $n\ [\times 10^{20}\mathrm{m}^{-3}]$             & 1.0                    &    2.0                         \\
            $\eta\ [\Omega \mathrm{m}]$                      & $1.93 \times 10^{-6}$  &  $3.49 \times 10^{-7}$   \\
            $\eta_{\mathrm{num}}\ [\Omega \mathrm{m}^3]$     & $1.93 \times 10^{-12}$ &  $3.49 \times 10^{-13}$     \\
            $\mu_\perp\ [\mathrm{kgm}^{-1}\mathrm{s}^{-1}]$                 & $5.15 \times 10^{-9}$  &  $6.46 \times 10^{-9}$    \\
            $\mu_{\mathrm{num}}\ [\mathrm{kgms}^{-1}]$       & $5.15 \times 10^{-15}$ &  $6.46 \times 10^{-15}$    \\
            $\mu_{\parallel,\ \bot}\ [\mathrm{kgm}^{-1}\mathrm{s}^{-1}]$                & -                      & $5.15 \times 10^{-7}$           \\
            $\mu_{\parallel,\ \parallel}\ [\mathrm{kgm}^{-1}\mathrm{s}^{-1}]$           & -                      & $5.15 \times 10^{-7}$           \\
            $D_\bot [\mathrm{m}^2\mathrm{s}^{-1}]$                    & $0.154$                & $0.154$                   \\
            $D_\parallel [\mathrm{m}^2\mathrm{s}^{-1}]$               & $1540$                 & -                         \\
            $\kappa_\bot [\mathrm{m}^2\mathrm{s}^{-1}]$               & $0.231$                & $0.231$                   \\
            $\kappa_\parallel [\mathrm{m}^2\mathrm{s}^{-1}]$          & $0.231\times 10^7$              & $4.52 \times 10^7$      \\
            $S_e\ [\mathrm{MW}]$                                    & $3.2$                  &   $3.2$                  \\ 
            $\mathrm{n}_\mathrm{rad}$                               & $65$                   &    $46$                 \\
            $\mathrm{n}_\mathrm{pol}$                               & $91$                   &    $64$                 \\
            $\mathrm{n}_\mathrm{tor}$                               & 0,5...35,40                   &    0,5,...45,50                
      \end{tabular}
      \label{tab:w7as_beta}
\end{table}

\begin{figure}
    \centering
    \includegraphics[width=0.5\textwidth]{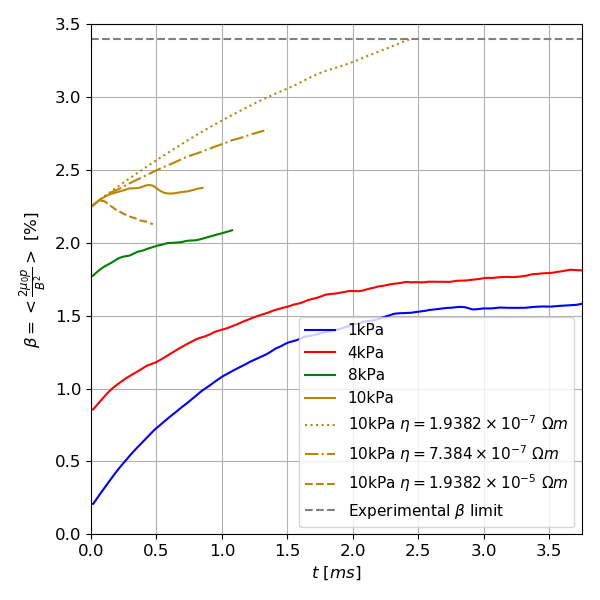}
    \caption{Evolution of $\beta$ for W7-AS cases from Figure \ref{fig:w7as_beta_prof}. Parameter scans of resistivity for the $10\ \mathrm{kPa}$ case show that the MHD activity is resistive.}
    \label{fig:w7as_beta_scan}
\end{figure}

The results for the evolution of the volume averaged plasma $\beta$ are shown in Figure \ref{fig:w7as_beta_scan}. The solid lines are run with the baseline parameters in Table \ref{tab:w7as_beta}. Considering first the $1\ \mathrm{kPa}$ and $4\ \mathrm{kPa}$ cases, the heat source is able to increase the plasma $\beta$ to $>1.3\ \%$ before the equilibrium approaches a flat top phase. The final $\beta$ is lower than the experimental value. The reason for this is that the simulations are fixed boundary, such that the induced Shafranov shift of the plasma is reduced. This means that the change in the $\iota$ profile observed in the free boundary equilibrium reconstructions in Figure \ref{fig:w7as_beta_prof} (a), which removes the $\iota=0.5$ rational surface from the system with increasing tokamak-like shear, is not reproduced in the simulation. In such a way, it is expected that the MHD activity is artificially enhanced by broader mode structures corresponding to the lower order $\iota=0.5$ rational surface that is not eliminated in the simulation, so that higher $\beta$ cannot be achieved.

Considering the baseline $10\ \mathrm{kPa}$ case, the volume averaged $\beta$ also initially rises, before a mild crash is observed, such that $\beta$ is limited to $\approx 2.35\ \%$. The nature of the crash is identified as resistive by conducting a parameter scan. Increasing the resistivity by an order of magnitude, it can be seen that the initial MHD activity is sufficient to cause a loss of confinement, such that there is an overall reduction in $\beta$. If the resistivity is reduced, the observed MHD activity becomes milder, and localised to the plasma edge, where the pressure gradient is largest. It is even shown that the experimental $\beta$ limit can be exceeded for $\eta = 1.93 \times 10^{-7}\ \Omega \mathrm{m}$.

\subsection{Simulations with experimental Spitzer resistivity} \label{sec:w7as_experimentally_relevant}
While the results in Section \ref{sec:w7as_beta_scan} are encouraging, a necessary next step is to confirm that a soft $\beta$ limit is still observed using more realistic parameters, which approach experimental relevance. In this Section, the Spitzer resistivity and Spitzer-Haerm conductivity are used with the appropriate temperature dependencies. As high $\beta$ operation of W7-AS was known to be in the turbulent transport regime, accurate calculations of the physical parameters for viscoresistive MHD simulations could not be found in the literature, beyond the approximate density of such discharges. For this reason, the experimental values for the viscosity and diffusion coefficients need to be approximated. As a recent study of RFPs shows, it is not easy to approximate the experimentally relevant fusion parameters for the viscoresistive MHD model, even when relying on significant experimental data\ \cite{vivenzi2022kinematic}.

Approximate values for W7-AS are shown in the righthand column of Table \ref{tab:w7as_beta}. The tabulated values correspond to the core of the device. The perpendicular transport coefficients are kept the same as the baseline scan in Section \ref{sec:w7as_beta_scan}. The perpendicular viscosity is chosen to be in the range of the expected Braginskii viscosity, with a Spitzer-like temperature dependence. The parallel viscosity coefficient is chosen to be constant and several orders of magnitude higher than the perpendicular viscosity, similar to what is typically expected from the Braginskii closure. The $10\ \mathrm{kPa}$ case is re-run with these parameters to determine whether the enhanced resistivity and parallel conductivity, as well as the inclusion of $\mathrm{v}_\parallel$ modify the dynamics.


\subsubsection{Numerical resolution scans}
\begin{figure}
    \centering
    \includegraphics[width=0.49\textwidth]{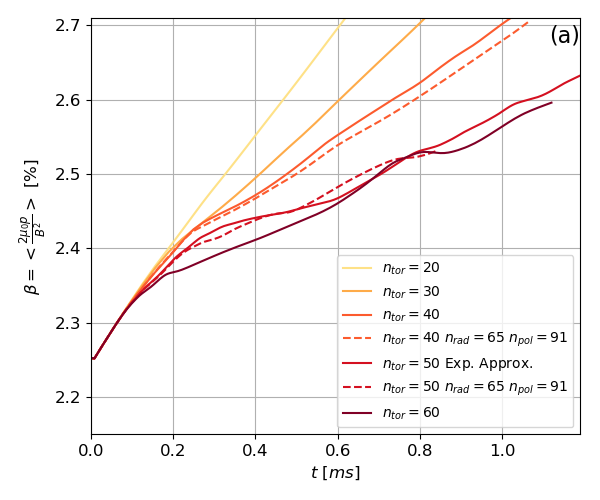}
    \includegraphics[width=0.49\textwidth]{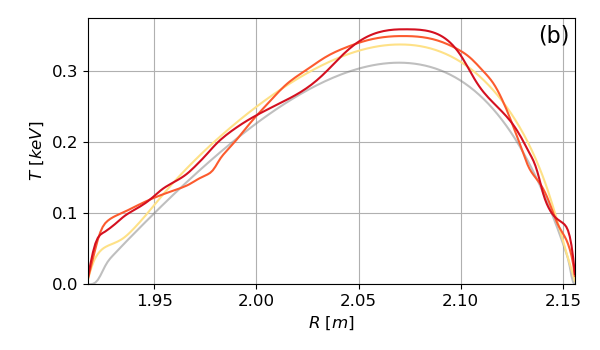}
    \caption{Evolution of $\beta$ (a) for the $10\ \mathrm{kPa}$ W7-AS case with different numerical resolution parameters. The toroidal resolution scans (solid lines) show that high $n$ modes are important in the dynamics, such that under-resolved simulations lead to artificially improved confinement, where MHD activity grows more slowly and saturates at a lower amplitude. Higher poloidal resolution does not lead to a significant change in the dynamics. The temperature of these cases in the $\phi=0$ midplane is plotted at the time point when $\beta \approx 2.55\ \%$ (b, red lines). The initial temperature profile is plotted for comparison (grey line). As $\mathrm{n}_\mathrm{tor}$ increases, the MHD activity is able to penetrate further into the core.}
    \label{fig:w7as_exp_beta_scan}
\end{figure}

In order to demonstrate the convergence of results, numerical resolution scans, particularly for the toroidal resolution, are first conducted. The change in plasma $\beta$ and temperature across the midplane in the $\phi=0$ poloidal plane when $\beta=2.55\ \%$ are shown in Figure \ref{fig:w7as_exp_beta_scan}. It can be seen that the initial the rate of increase of $\beta$ over time is strongly influenced by the toroidal resolution, decreasing as the toroidal resolution is increased up to  $\mathrm{n}_\mathrm{tor} \approx 50$. 

This implies that high $n$ modes play an important role in the stochastisation, and subsequent enhanced transport of the initial equilibrium. The initial MHD activity grows faster with more toroidal harmonics, and saturates at a higher amplitude (not shown). The temperature profiles in Figure \ref{fig:w7as_beta_scan} (b) show that this increased energy corresponds to a deeper penetration of the MHD perturbation into the plasma core. Consequently, the temperature profile becomes more peaked for the same plasma $\beta$, with local regions of stronger ergodisation where the temperature profile is partially flattened. 

For  $\mathrm{n}_\mathrm{tor} \ge 50$, the rate of increase in $\beta$ does not vary significantly. In the  $\mathrm{n}_\mathrm{tor}=50$ and  $\mathrm{n}_\mathrm{tor}=60$ cases, there are brief periods where more significant stochastisation occurs, around $t=0.4\ \mathrm{ms}$ and $t=0.83\ \mathrm{ms}$, respectively. As discussed in more detail in Section \ref{sec:w7as_evid}, this corresponds to an overlap of island structures in the plasma mid-region. The plasma soon recovers in both cases, such that $\beta$ continues to rise at approximately the same rate in both simulations. Comparing the $\mathrm{n}_\mathrm{tor}=40$ and $\mathrm{n}_\mathrm{tor}=50$ case with similar runs using a higher poloidal resolution, it can be seen that the dynamics are well resolved with respect to the poloidal resolution parameters.

\subsubsection{Conductivity parameter scans}
In order to test how important the prescribed conductivity parameters are in defining the dynamics, scans of the perpendicular and parallel conductivity coefficients are also carried out in Figure \ref{fig:w7as_beta_scan}. The parameter scans of the conductivity are used to demonstrate the importance of the prescribed conductivity parameters in such MHD simulations. The blue line shows a case where the perpendicular conductivity is increased. As expected, this reduces the initial rate of increase in $\beta$ due to worse confinement. As the stochasation of the plasma occurs later in time, $\beta$ begins to decrease, leading to a reduction from the initial equilibrium thermal energy. In this sense, confinement is at least partially lost.

\begin{figure}
    \centering
    \includegraphics[width=0.49\textwidth]{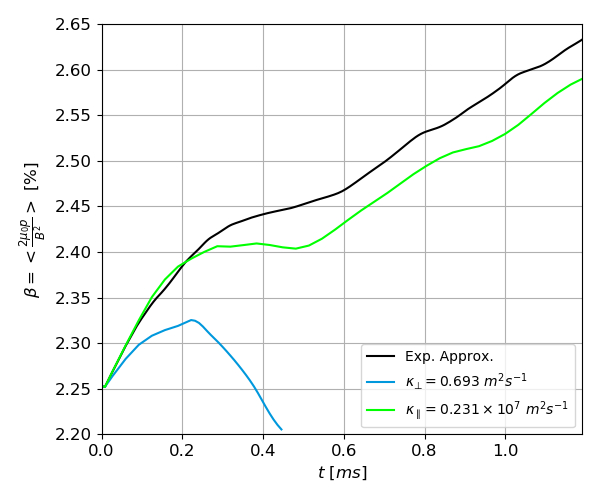}
    \caption{Modifications to the conductivity (blue and green lines) show that these prescribed parameters are important for predicting the evolution of $\beta$.}
    \label{fig:w7as_exp_beta_physics_scan}
\end{figure}

The green line in Figure \ref{fig:w7as_exp_beta_scan} shows a case where the parallel conductivity is assumed to be constant, and reasonably large across the plasma volume. This means that the conductivity is larger than in the experimental approximation in the edge region where the MHD activity initially develops. As the parallel conductivity stabilises pressure driven MHD activity \cite{xu2011nonlinear}, $\beta$ initially grows slightly more quickly than with the baseline parameters, due to a delayed onset of the ergodisation of field lines. Later in time, when the edge region becomes stochastic, the rate of increase in $\beta$ is reduced, but continues to grow at a comparable rate to the experimental approximation. The results of these conductivity scans imply that the soft $\beta$ limit result in Figure \ref{fig:w7as_exp_beta_scan} (a) is sensitive to the thermal transport parameters.

\subsubsection{Ergodic plasma confinement} \label{sec:w7as_evid}
Using the experimental approximation with the parameters shown in the right column of Table \ref{tab:w7as_beta}, the thermal energy remains sufficiently well confined outside the bounds of linear MHD stability, to reach higher $\beta$, even after the nonlinear saturation of the MHD activity. It is expected that a soft $\beta$ limit will eventually be observed when the transport channels are large enough to balance the input heating power.

\begin{figure*}
    \centering
    \includegraphics[width=\textwidth]{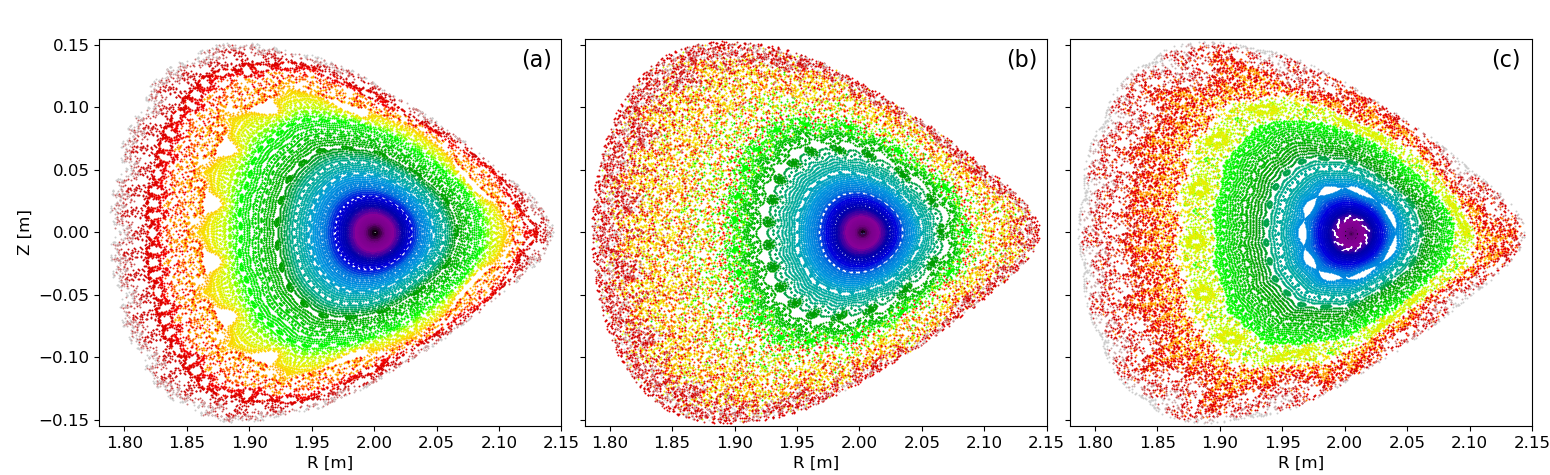}
    \caption{Poincar\'e plots of the simulation case with the baseline parameter in the righthand column of Table \ref{tab:w7as_beta} (red line, labelled $\mathrm{n}_\mathrm{tor}=50$ in Figure \ref{fig:w7as_exp_beta_scan} (a)). The plots are taken at the initial saturation (a, $t=0.188\ \mathrm{ms}$), the period of worst confinement (b, $t=0.421\ \mathrm{ms}$), and near the end of the simulation time (c, $t=1.189\ \mathrm{ms}$). Though the degree of ergodisation varies over the simulation time, the plasma remains separated into distinct radially localised sub-volumes.}
    \label{fig:w7as_exp_beta_poincare}
\end{figure*}

Figure \ref{fig:w7as_exp_beta_poincare} shows Poincar\'e plots for the $\mathrm{n}_\mathrm{tor}=50$ case at three time points corresponding to the initial saturation, period of worst confinement, and end of the simulation time. It can be seen that while the initial flux surfaces have become ergodic in Figure \ref{fig:w7as_exp_beta_poincare} (a), the field lines can still be separated into localised regions of the total plasma volume. Individual island structures can be identified, some of which correspond to relatively high poloidal mode numbers. This gives an indication that the MHD induced island structures are not large enough to significantly overlap, which would have flattened the temperature profile over a broader region. In such a way, the enhanced perpendicular transport through heat transfer along the stochastic magnetic field is mild and $\beta$ can continue to increase. 

Following the initial saturation, the magnetic energy continues to fluctuate with periods of lower and higher MHD activity. There is a brief period where the internal ergodic regions merge transiently, as shown in Figure \ref{fig:w7as_exp_beta_poincare} (b). In particular, this enhanced transport occurs when the ergodic region penetrates beyond the $\iota=0.5$ rational surface, which can be identified in  Figure \ref{fig:w7as_exp_beta_poincare} (a) and (c) by the corresponding (10, 5) magnetic island chain. Even so, during this period, several sub-volumes can still be identified, such that confinement has not been completely lost. Later in time, the plasma recovers from this transient phase, and good confinement is preserved in the plasma core. For most of the simulated time, the main ergodisation is observed in the outer region of the plasma, outside the $\iota={0.5}$ rational surface. This observation is similar to that of past equilibrium studies for W7-AS using PIES, where the stochastic region was predominantly in the outer region of the plasma volume \cite{Reiman2007}.

\subsection{Observation of low $n$ mode activity} \label{sec:w7as_low_n_activity}

Full torus simulations have been carried out for the $10\ \mathrm{kPa}$ case to test whether low $n$ modes are nonlinearly dominant, as observed in the experiment in the intermediate $\beta$ range, $1.5 \le \beta \le 2.5$. Two simulations were run using both the simplified parameters in the central column of Table \ref{tab:w7as_beta} with an artificially high resistivity, $\eta=1.93 \times 10^{-5}\ \Omega \mathrm{m}$, and the experimental Spitzer resistivity with the parameters in the righthand column of Table \ref{tab:w7as_beta}. For this initial study, a lower toroidal resolution, $\mathrm{n}_\mathrm{tor}=20$ is used to make simulations more computationally tractable. 

\begin{figure}
    \centering
    \includegraphics[width=0.5\textwidth]{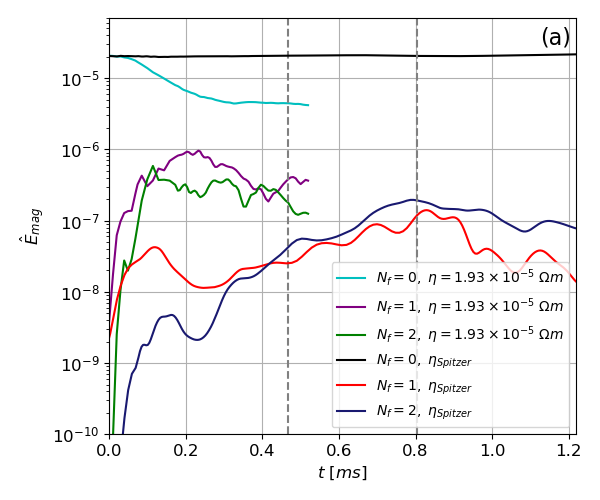}
    \includegraphics[width=0.5\textwidth]{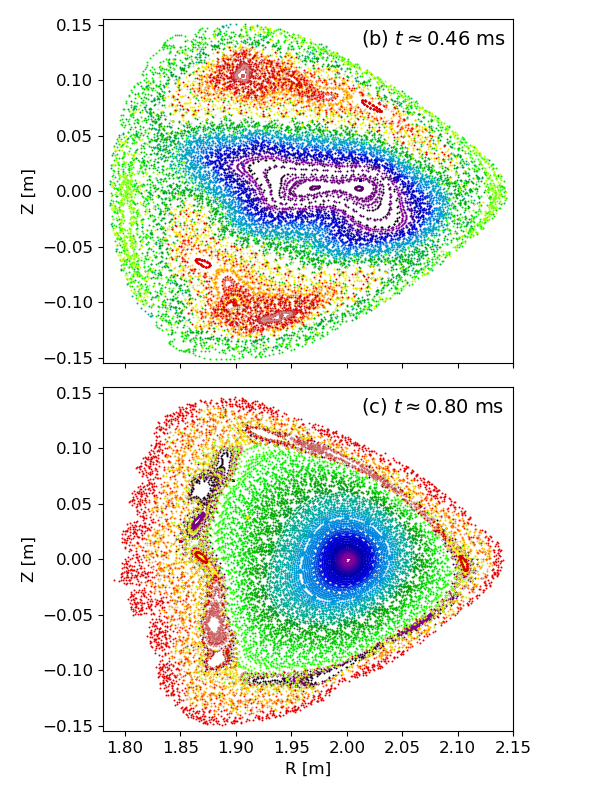}
    \caption{Normalised magnetic energy traces (a) of full torus W7-AS simulations with artificially high resistivity (cyan, green, and purple lines) and the Spitzer resistivity (black, blue and red lines). Low $n$ modes are dominant over most of the nonlinear phase in both cases. The $n=1$ harmonic corresponds to the (2, 1) mode. A Poincar\'e plot for the case with $\eta = 1.93 \times 10^{-5}\ \Omega \mathrm{m}$ (b) taken at $t \approx 0.46\ \mathrm{ms}$ shows the dominant (2, 1) island structure in the mid-region of the plasma. The Poincar\'e plot in the case using the Spitzer resistivity (c), taken at $t \approx 0.80\ \mathrm{ms}$, has a less obvious mode structure, with a (4, 2) structure interacting with higher $n$ modes at the $\iota=0.5$ rational surface.}
    \label{fig:w7as_full_torus}
\end{figure}

The normalised magnetic energy of each toroidal mode family is shown in Figure \ref{fig:w7as_full_torus} (a). It can be seen that in both simulations, the  $N_\mathrm{f}=1$ mode family saturates first, though for the high resistivity case, the magnetic perturbation is larger, indicating again that the MHD activity is resistive. For the high resistivity case, the saturation seems large enough to affect the  $N_\mathrm{f}=0$ mode family significantly, and the initial thermal energy stored in the plasma is gradually lost, similar to the single field period case in Figure \ref{fig:w7as_beta_scan}.  For the experimental resistivity, good confinement is preserved sufficiently for the thermal energy to continue to increase when the experimental heat source is applied.

In both cases, the  $N_\mathrm{f}=2$ mode family is observed to compete with the  $N_\mathrm{f}=1$ modes. For the high resistivity case, the  $N_\mathrm{f}=1$ mode remains dominant over most of the nonlinear phase. A Poincar\'e plot taken near the end of the simulation time in Figure \ref{fig:w7as_full_torus} (b) shows the nonlinearly dominant (2, 1) mode. A (4, 2) sub-structure remains visible, showing the competition between the two mode families. The observation of this nonlinearly dominant (2, 1) mode is in line with past experimental observations. However, it should be noted that, at this higher resistivity, the (2, 1) perturbation is very large at earlier time points in the simulation, spreading over most of the plasma volume.

In the case with the experimental Spitzer resistivity, the energies in Figure \ref{fig:w7as_full_torus} (a) show that the $N_\mathrm{f}=2$ mode family dominates in the nonlinear phase. The $N_f=0$ energy also remains large, however. Considering the Poincar\'e plot in Figure \ref{fig:w7as_full_torus} (c), good flux surfaces are preserved within the plasma core in this case. A (4, 2) deformation of the flux surfaces is discernible near the expected location of the $\iota=0.5$ rational surface, however the island structures show sub-structures corresponding to higher poloidal mode numbers. This indiates that the $N_\mathrm{f}=0$ modes still play a role in the dynamics, making the plasma ergodic, such that distinct low $m$ island structures are difficult to identify around the low order rational surface.

\subsection{Discussion}\label{sec:w7as_discussion}
The analysis in this section has made a first pass at understanding the soft $\beta$ limits in W7-AS. Regarding the three initial goals defined at the beginning of the Section, the simulations support the conclusion that the MHD activity is viscoresistive. At viscoresistive parameters which approach the experimental values, a soft $\beta$ limit can be observed, because the induced MHD activity is mild enough for field lines to remain ergodically confined within nested plasma sub-volumes. When the full torus is modeled, low $n$ modes play an important role in the MHD activity in the nonlinear phase. This supports the conclusion that JOREK can correctly reproduce the experimental findings.


While these initial results are encouraging, there are limitations to the current modelling capabilities. If the numerical model were to be comparable to the experiment, the simulated plasma would eventually reach a flat top phase, where transport channels would counter-balance the input heating power. This would be expected around the experimental $\beta$ limit, $3.4\ \%$. There are several contributing factors which could explain why this is not observed in the simulations in Section \ref{sec:w7as_beta_scan} and \ref{sec:w7as_experimentally_relevant}. 

Firstly, the fixed boundary condition means that while anomalous perpendicular transport due to parallel transport along magnetic field lines can arise, field lines are enforced to be parallel to the simulation boundary. This is a poor approximation of the island divertor used in the experiment, where parallel heat conduction is expected to play an important role in radial transport in the scrape off layer. While the connection length of an island divertor is still significant \cite{feng2006physics}, such that perpendicular dynamics will still play a role in the edge transport, the current boundary conditions cannot be expected to fully reproduce the thermal losses of the experiment.

On the other hand, as discussed in Section \ref{sec:w7as_beta_scan}, the fixed boundary will also reduce the increase in the Shafranov shift, and tokamak-like shear that is developed in the experiment. This is an artificial destabilising effect in the simulation, as the low order $\iota=0.5$ rational surface is not removed from the system with increased $\beta$. Further work to include the external divertor region, and introduce a free boundary condition on the magnetic field would be necessary to achieve this level of fidelity. These numerical developments are beyond the scope of the current work.

In addition to the above limitations of the MHD modeling, it should be noted that the transport coefficients used herein, are approximate. As shown in Section \ref{sec:w7as_experimentally_relevant}, simulations with larger perpendicular heat conductivity can lead to a loss in confinement. High $\beta$ W7-AS operation was in the turbulent transport regime. Thus, detailed calculations of the transport coefficients using turbulence codes would be necessary to get this aspect of the transport right, which has not been attempted as part of this study. Furthermore, the influence of impurities and radiative losses, which are known to play a role in W7-AS, are not included in the modeling. While soft radiative core collapse events could be avoided after divertor installation \cite{giannone2002radiation}, the neglected radiative losses could influence the power balance in simulations.

Regarding the observation of the (2, 1) resistive ballooning mode at intermediate $\beta$, an artificially high resistivity was used to observe this mode. For the simulation with more realistic parameters, nonlinearly dominant low $n$ modes are still observed, but the saturated modes have significantly lower energy such that the corresponding island structures are not so easily distinguished from higher $n$ modes in Poincar\'e diagnostics. There are several reasons for this. Firstly, the toroidal resolution is potentially too low, which may have reduced the nonlinearly saturated energy of the perturbation, similar to Section \ref{sec:w7as_experimentally_relevant}. Furthermore, the fixed boundary conditions, which will preferentially stabilise the low $n$ modes, may have prevented the (2, 1) mode from being dominant. 

Finally, the neglect of the ambipolar radial electric field in the current study, which introduces a sheared $\mathbf{E} \times \mathbf{B}$ flow into the system, could have influenced the dynamics. Such a sheared flow could stabilise high $n$ modes, while exacerbating low $n$ modes in a similar way to what is observed for edge harmonic oscillations \cite{xu2017b}, which the low $n$ MHD activity on W7-AS has been compared to \cite{hirsch2008major}. This effect could allow the $n=1$ dominant mode to lead the dynamics. It should be noted that the (2, 1) mode observed experimentally on W7-AS was shown to be poloidally rotating \cite{weller2003experiments}, such that flows are likely to be an important aspect of the dynamics, which has thus far been neglected. 

Acknowledging these limitations of the current modeling, considerable progress has nevertheless been made towards a simulation set up that captures many of the physical features of the experiment, such that it is a reasonable approximation of the nonlinear dynamics. The results so far provide a strong indication that soft $\beta$ limits can be observed in JOREK, in agreement with the experiment. The major unanswered question for the community is the reason that the low $n$ MHD activity saturates at mild values. The results in Section \ref{sec:w7as_experimentally_relevant} indicate that the ergodisation introduced by the saturated MHD activity is mild as there is not a large overlap of magnetic island structures across the plasma volume. Understanding why and under what conditions the saturated island structures are sufficiently small to observe a soft $\beta$ limit is the subject of ongoing work, which needs to be coupled to ongoing performance work and extensions to the MHD model, in order to make realistic full torus simulations computationally tractable.

\section{Conclusion} \label{sec:conclusion}
Several important steps have been taken towards building a credible case for the study of $\beta$ limits in stellarators using JOREK in the current work. Firstly, the model in JOREK has been extended to capture the influence of parallel flows and fluid compressibility. Verification studies using the model to simulate strongly shaped geometries, and finite $\beta$ equilibria show that excellent agreement can be obtained over most of the cases considered, particularly in the low magnetic shear regime. This indicates that there are finite $\beta$ cases with low magnetic shear, which can be studied with the reduced MHD model in JOREK. Where there is disagreement, reasonable explanations to do with the simulation set up, namely the elimination of external modes, are put forward. It is further shown that these eliminated modes can be observed in JOREK, but require further free boundary extensions to benchmark properly.

Turning to studies of experiments, a first pass has been made to simulate the soft $\beta$ limits observed on W7-AS. Initial simulations show that the MHD activity observed is viscoresistive, and the observed soft $\beta$ limits strongly depend on these parameters. In the regime where an experimentally relevant resistivity is used, a soft $\beta$ limit is observed. This seems to be due to the mild overlap of the magnetic island structures induced by MHD activity, which is not able to significantly enhance the perpendicular transport, and flatten the temperature profile, because magnetic field lines remain ergodically confined. Further work, incorporating the divertor, $\mathbf{E} \times \mathbf{B}$ flows, and free boundary conditions would improve the fidelity of the modelling of low $n$ modes. It is intended to carry out these extensions before making a second pass at this case, in an attempt to explain the suppression mechanism of the (2, 1) resistive ballooning mode observed in this study.

Beyond W7-AS, there is of course interest in building towards studies of fully optimised stellarators, and understanding their $\beta$ limits. For most devices, including W7-X, this requires predictive simulation studies, as the $\beta$ limit has not yet been studied experimentally in detail. This is one of the main reasons for focusing on W7-AS and more extensively characterised stellarators in the near term --- so that a strong validation can be used to justify the use of the numerical methods herein predictively in the coming years. It is possible to study electron temperature crashes at finite $\beta$ in W7-X, and initial studies have been developed towards this goal. Initial preparatory studies of pressure driven MHD activity in W7-X, as well as MHD activity in quasi-symmetric stellarator concepts are in progress towards the longer term goal of predictive studies. 

\section*{Acknowledgements}
The authors would like to thank Nicola Isernia, Andres Cathey, Jonas Puchmayr, Guido Huijsmans, Karl Lackner and Sibylle G\"unter for helpful discussions. In addition, the authors are grateful to Carolin Nuehrenberg and Joachim Geiger for providing some of the equilibria necessary for the studies herein, as well as helpful discussions and insights, in particular, regarding W7-AS high $\beta$ operation. 

Some of this work was carried out on the high performance computing architectures COBRA and RAVEN operated by MPCDF in Germany, JFRS-1 operated by IFERC-CSC in Japan, and the EUROfusion High Performance Computer (Marconi-Fusion).

This work has been carried out within the framework of the EUROfusion Consortium, funded by the European Union via the Euratom Research and Training Programme (Grant Agreement No 101052200 — EUROfusion). Views and opinions expressed are however those of the author(s) only and do not necessarily reflect those of the European Union or the European Commission. Neither the European Union nor the European Commission can be held responsible for them.

\section*{References}
\bibliographystyle{iopart-num}
\bibliography{references}

\end{document}